# Lagrangian Modelling and Motion Stability of Synchronous Generator-based Power Systems

Feng Ji, *Member, CSEE*, Lu Gao, Chang Lin, *Member, CSEE*, Yang Liu, *Member, CSEE*

*Abstract*—This paper proposes to analyze the motion stability of synchronous generator-based power systems using a Lagrangian model derived in the configuration space of generalized position and speed. A Lagrangian model of synchronous generators is derived based on Lagrangian mechanics. The generalized potential energy of inductors and generalized kinetic energy of capacitors are defined. The mechanical and electrical dynamics can be modelled in a unified manner through constructing a Lagrangian function. Taking the first benchmark model of sub-synchronous oscillation as an example, a Lagragian model is constructed and numerical solution of the model is obtained to validate the accuracy and effectiveness of the model. Compared with the traditional EMTP model in PSCAD, the obtained Lagrangian model is able to accurately describe the electromagnetic transient process of the system. Moreover, the Lagrangian model is analytical, which enables to analyze the motion stability of the system using Lyapunov's motion stability theory. The Lagrangian model can not only be used for discussing the power angle stability, but also for analyzing the stability of node voltages and system frequency. It provides the feasibility for studying the unified stability of power systems.

*Index Terms*—motion stability, Lagrangian mechanics, Lyapunov's first method, sub-synchronous resonance.

## I. Introduction

Power system is a complex dynamic system and its stability is an inherent requirement of energy consumers. For more than a hundred years, the AC power system has been built on the synchronous operation mechanism of synchronous generators. However, with the development of the high proportion renewable energy and power electronics, the fundamental modelling and analyzing theory of power system stability is facing unprecedented challenges [1]. On one hand, the time-varying output of renewable power leads to the rapid migration of system operating points. The traditional stability theory based on given equilibrium points would not be adaptive [2]. On the other hand, electromagnetic resonant problems are so common that dynamics of inductors and capacitors cannot be omitted in system modelling anymore. In order to cope with these problems, we need to backtrack the modelling and analyzing history of AC power systems, and make fundamental developments to adaptive to the new characteristics of modern power systems.

In 1892, Lyapunov developed a complete theory to analyze the motion stability of ordinary differential equations $d\mathbf{x}/dt = \mathbf{f}(\mathbf{x}, t)$ [3][4]. Lyapunov's theory has been widely used for analyzing the stability of power systems modelled by the electromechanical transient model [5]. Power angle stability, voltage stability, and frequency stability were analyzed on the basis of the model [6]. In the electromechanical transient model, rotor angles of generator are state variables, which are kept at the synchronization position during steady-state operation [7]. The electric network is modelled in the form of algebraic equations. The node voltages and frequency in the electric network are not modelled as state variables. Actually, there are no precise definitions for voltage stability and frequency stability, although many international seminars have been held, and even IEEE/CIGRE published corresponding special reports [6].

Section 12.3.1 in [7] gives a proof for the power angle stability of the general single-machine infinite-bus (SMIB) system, based on its electromechanical transient model and Lyapunov's first method. However, a SMIB system is not always stable in terms of power angle. For example, the first benchmark model of sub-synchronous oscillation is a typical unstable SMIB system [8]. Instability of this model can only be reflected and analyzed using electro-magnetic transient analysis programs (EMTPs). Accounting for the electromagnetic dynamics-dominated power electronics devices, it is believed that the electromechanical transient model is no longer suitable for the dynamic characteristic analysis of the power system with high proportion of power electronics and renewable generation [9]. EMTPs based on detailed dynamic models of components are able to accurately describe the dynamic process of the system at microsecond level, and has gradually been regarded as an effective and accurate simulation method for power electronic power system [10][11].

Nevertheless, power networks in EMTPs are modeled by algebraic equations in the form of $\mathbf{Gu}=\mathbf{i}$ [12][13] rather than a differential equation, where $\mathbf{G}$ is the admittance matrix, $\mathbf{u}$ is the node voltage vector, and $\mathbf{i}$ is the vector of the injected current of nodes. Hence, the EMTP model cannot reveal the dynamics of power networks, and analytical analysis of system stability cannot be carried out as well. An ordinary differential equation (ODE) model was derived for single-phase RLC circuits in [14] to describe the electromagnetic transients based on the variational principle. [15] proposed the ODE model of three-phase RLC circuits, and also derived the model in the synchronously rotating coordinate system. On the basis of the above works, this paper derives an ODE model for synchronous generator-based power systems via Lagrangian mechanics, and proposes to analyze the motion stability of power systems using the Lagrangian ODE model.

In order to illustrate the distinctions and superiorities of the proposed Lagrangian model in comparison to the model embedded in a typical EMTP, i.e., PSCAD, this paper takes the first benchmark model of sub-synchronous oscillation as an example. By selecting the power angle of the synchronous generator and the node flux linkages of circuit nodes as the generalized positions, the Lagrangian model of the benchmark system is built. Numerical simulation results of the Lagrangian model validate that Lagrangian ODE model is able to accurately describe the electromagnetic transient processes of the benchmark system. Moreover, analytical calculation of oscillation modes of the benchmark system is carried out, and highly agreements are achieved between the analytical results and simulation results. Furthermore, the node voltages and frequency in the electric network are modeled as the state variables of the Lagrangian ODE model, thereby providing a feasibility for accurately defining and analyzing the voltage stability and frequency stability.

Overall, this paper is organized as follows. The knowledge of Lyapunov's motion stability and the first stability criterion is introduced in Section II. The Lagrangian Modelling of the synchronous generator is undertaken in Section III, in which the first benchmark model of sub-synchronous oscillation is modelled with Lagrangian mechanics. Formulation of the stability problem of the Lagrangian model in sense of equilibrium is carried out in Section IV. Numerical validation is presented in Section V. The Lagrangian modelling is compared with other circuit modelling methods in Section VI, and conclusions follow thereafter in Section VII.

## II. LYAPUNOV'S MOTION STABILITY

The original definition of motion stability given by Lyapunov is very complex. This paper adopts the simplified definition of motion stability [16], which was given by Tsien Hsue-shen.

$$\frac{d\mathbf{x}}{dt} = \mathbf{f}(\mathbf{x},t), \quad \mathbf{x} = \begin{bmatrix} x_1 & x_2 & \cdots & x_n \end{bmatrix}^T \quad (1)$$

For a system described by (1), suppose that $\mathbf{x}(t)$ is a solution of (1) when the initial condition is $\mathbf{x}(t_0) = \mathbf{x}_0$. The definition of the stability of this solution is as follows: given an arbitrary positive number $\varepsilon > 0$, there is always a number $\delta$ greater than zero, when the disturbance of the initial condition $\mathbf{x}_0$ at time $t = t_0$ is less than $\delta$ ($\|\boldsymbol{\delta x}_0\| < \delta$), the change of the corresponding solution of (1) is always less than $\varepsilon$ at any time of $t \geq t_0$ (as shown in Fig.1a), then this particular solution of (1) is stable. It can be seen that the notion of motion stability is different from the commonly used notion of stability in terms of equilibriums.

In order to apply the Lyapunov's first stability criterion, a coordinate transformation should be made on the original equation. Let $\mathbf{x}(t)$ be the motion with the initial condition $\mathbf{x}_0$, and the disturbed motion $\mathbf{z}(t)$ be the motion corresponding to the initial condition $\mathbf{x}_0 + \Delta \mathbf{x}_0$. Making $\Delta \mathbf{x}(t) = \mathbf{z}(t) - \mathbf{x}(t)$, it satisfies (2).

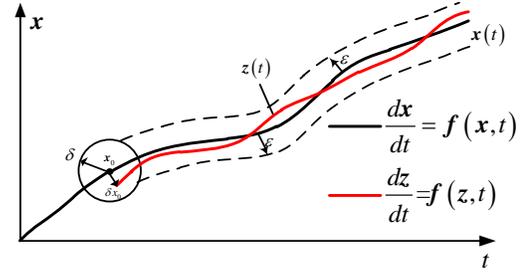

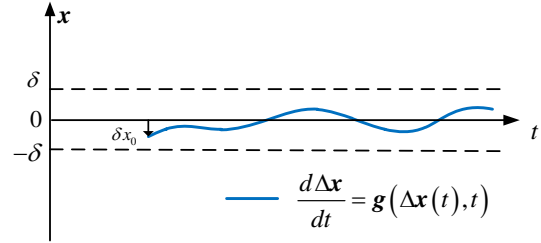

Fig. 1. Schematic of motion stability.

$$\begin{aligned} \frac{d\mathbf{z}}{dt} - \frac{d\mathbf{x}}{dt} &= \mathbf{f}(\mathbf{z},t) - \mathbf{f}(\mathbf{x},t) \\ \frac{d\Delta\mathbf{x}}{dt} &= \mathbf{f}(\mathbf{x}(t) + \Delta\mathbf{x}(t), t) - \mathbf{f}(\mathbf{x},t) \quad (2) \\ \frac{d\Delta\mathbf{x}}{dt} &= \mathbf{g}(\Delta\mathbf{x}(t), t) \end{aligned}$$

where $\Delta\mathbf{x}(t_0) = \Delta\mathbf{x}_0$. Obviously $\Delta\mathbf{x}(t) \equiv 0$, when $\Delta\mathbf{x}_0 = 0$. In this way, the motion stability problem of (1) is transformed into the stability of the zero solution of (2), as shown in Fig.1b.

Suppose that $\mathbf{f}(\mathbf{x},t)$ is a continuously differentiable function when the components of $\Delta\mathbf{x}(t)$ are small enough, and the right end of (2) can be expanded into (3) by Taylor expansion.

$$\frac{d\Delta\mathbf{x}(t)}{dt} = \mathbf{A}(t)\Delta\mathbf{x}(t) + \mathbf{h}(\Delta\mathbf{x}(t),t) \quad (3)$$

where $\mathbf{h}(\Delta\mathbf{x}(t),t)$ is the high-order remainder term. Lyapunov's first stability criterion can be reiterated as that system (3) is asymptotically stable, when the real part of all eigenvalues of $\mathbf{A}$ is negative; it is unstable, when $\mathbf{A}$ has one or more eigenvalue whose real part is positive. Accordingly, using the first stability criterion to determine the stability of a particular motion of a differential equation involves the follow steps:

Step 1: find the particular solution $\mathbf{x}(t)$ of the equation (1). If the particular solution does not exist, it is unstable;

Step 2: take $\Delta\mathbf{x}(t) = \mathbf{z}(t) - \mathbf{x}(t)$ as the difference between the disturbed motion and the particular solution, and obtain the differential equation $d\Delta\mathbf{x}/dt = \mathbf{g}(\Delta\mathbf{x}(t),t)$;

Step 3: judge the stability of the zero solution of $d\Delta\mathbf{x}/dt = \mathbf{g}(\Delta\mathbf{x}(t),t)$.

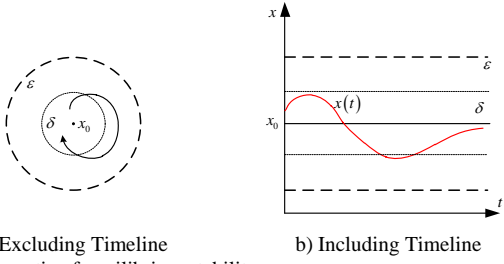

a) Excluding Timeline      b) Including Timeline
Fig. 2 Schematic of equilibrium stability.

If the particular solution $x(t)=x_0$ is a constant vector, the stability of motion degenerates into an equilibrium point stability problem, where $x_0$ is the equilibrium point. As shown in Fig.2, equilibrium stability is a special type of motion stability problem.

## III. LAGRANGIAN MODELLING OF SYNCHRONOUS GENERATORS

Taking the first benchmark model of sub-synchronous resonance as an example, this section uses the variational principle of Lagrangian mechanics to obtain the dynamic model of the synchronous generator. For the sake of simplicity, this paper removes the transformer from the original benchmark model, and converted the parameters of the line resistance, inductance and series compensation capacitance to the generator side, as shown in Fig.3.

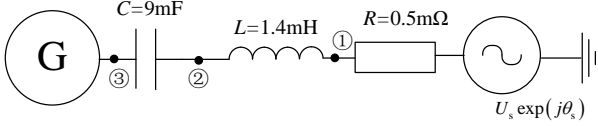

(a) Circuit diagram of the first benchmark model of sub-synchronous resonance.

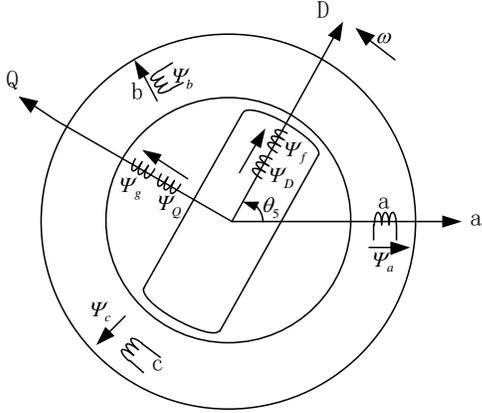

(b) Schematic diagram of a seven-winding generator

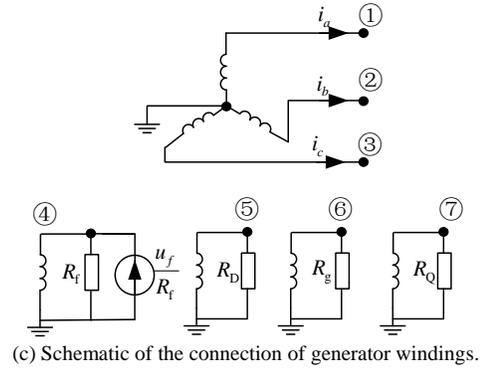

(c) Schematic of the connection of generator windings.

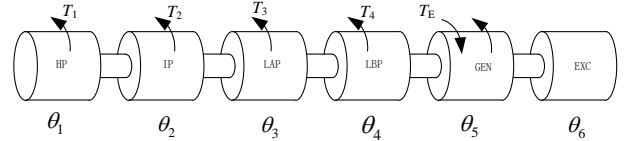

(d) Multi-mass model of the shafting system of the synchronous generator.

Fig. 3 Single machine infinite bus system with sub-synchronous oscillation.

In the first place, the voltage source representing the infinite-bus in Fig.3a should to be transformed into its Norton equivalent, as shown in Fig.4. Because the stator winding of the generator is connected to node 3, the flux of node 3 is the flux of the stator windings of the generator. In addition, the flux of the excitation winding f, the damping winding D, g, and Q are also the state variables of the system. There are (3×3+4=) 13 node flux to be considered in the modelling of the system in Fig. 3. Concerning a three-phase balanced case, the zero-axis components of the three-phase nodes can be eliminated by introducing the Clark transformation ignoring the zero-axis component. In this way, the number of the flux leakages to be modelled becomes (3×2+4=) 10.

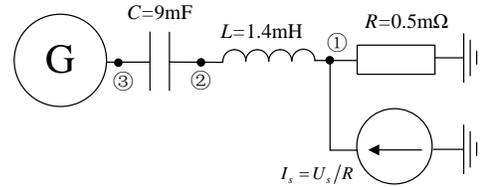

Fig. 4 Equivalent circuit of the benchmark model with Norton current source.

Hence, winding dynamics are modelled in the space of node flux $\boldsymbol{\Psi}=(\Psi_{1\alpha},\Psi_{1\beta},\Psi_{2\alpha},\Psi_{2\beta},\Psi_{3\alpha},\Psi_{3\beta},\Psi_f,\Psi_D,\Psi_g,\Psi_Q)^T$, whose derivative is $\dot{\boldsymbol{\Psi}}=(u_{1\alpha},u_{1\beta},u_{2\alpha},u_{2\beta},u_{3\alpha},u_{3\beta},u_f,u_D,u_g,u_Q)^T$. Meanwhile, the mechanical shafting dynamics are modelled in the space of angular position displacements, i.e., $\boldsymbol{\theta}=(\theta_1,\theta_2,\theta_3,\theta_4,\theta_5,\theta_6)^T$, whose derivative is the rotational angular velocity denoted by $\dot{\boldsymbol{\theta}}=(\dot\theta_1,\dot\theta_2,\dot\theta_3,\dot\theta_4,\dot\theta_5,\dot\theta_6)^T$.

### A. Definition of Potential Energy

According to [17], potential energy is a function of displacement variables. In Fig.3, the energy stored in the inductors and the windings of the generator is a function of node flux. The elastic potential energy in mechanical shafting is the function of angular displacements. The inductance branches in Fig.3 can be redrawn as in Fig 5, in which Fig.5a shows three single-phase inductance branches, and Fig.5b presents a three-phase inductance branch.

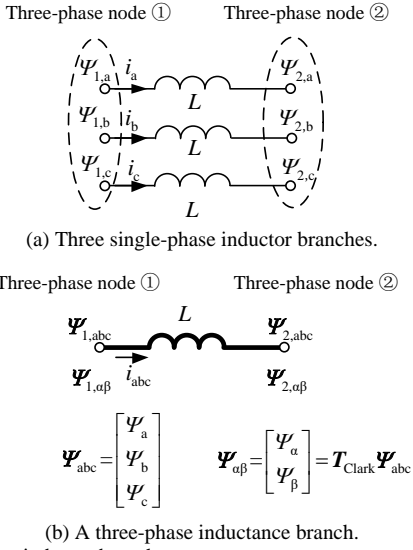

(a) Three single-phase inductor branches.

(b) A three-phase inductance branch.
Fig. 5 Three-phase inductor branches.

The energy stored in the three-phase inductor is

$$E_l = \frac{1}{2}Li_a^2 + \frac{1}{2}Li_b^2 + \frac{1}{2}Li_c^2 \tag{4}$$

Expressing branch currents with node flux, it has

$$\begin{cases} i_a = L^{-1}(\Psi_{1a} - \Psi_{2a}) \\ i_b = L^{-1}(\Psi_{1b} - \Psi_{2b}) \\ i_c = L^{-1}(\Psi_{1c} - \Psi_{2c}) \end{cases} \tag{5}$$

then (4) can be rewritten as

$$E_l = \frac{1}{2}L^{-1}(\Psi_{1a} - \Psi_{2a})^2 + \frac{1}{2}L^{-1}(\Psi_{1b} - \Psi_{2b})^2 + \frac{1}{2}L^{-1}(\Psi_{1c} - \Psi_{2c})^2 \tag{6}$$

where $\Psi_{1a}$, $\Psi_{1b}$ and $\Psi_{1c}$ are the node flux of phase a, b and c of node 1, respectively, $\Psi_{2a}$, $\Psi_{2b}$ and $\Psi_{2c}$ are the node flux of phase a, b, and c of node 2, respectively. Denote the three-phase node flux as $\Psi_{1,abc} = [\Psi_{1a}, \Psi_{1b}, \Psi_{1c}]^T$, $\Psi_{2,abc} = [\Psi_{2a}, \Psi_{2b}, \Psi_{2c}]^T$, and its Clark transformation is $\Psi_{1,\alpha\beta} = T_{\text{Clark}}\Psi_{1,abc}$ and $\Psi_{2,\alpha\beta} = T_{\text{Clark}}\Psi_{2,abc}$, where $T_{\text{Clark}} = \sqrt{\frac{2}{3}}\begin{bmatrix} 1 & -1/2 & -1/2 \\ 0 & \sqrt{3}/2 & -\sqrt{3}/2 \end{bmatrix}$ [18]. Equation (6) could be rewritten in the form of inner product of vectors as follows.

$$E_l = \frac{1}{2}L^{-1}(\Psi_{1,abc} - \Psi_{2,abc}, \Psi_{1,abc} - \Psi_{2,abc})$$
$$= \frac{1}{2}L^{-1}(\Psi_{1,\alpha\beta} - \Psi_{2,\alpha\beta}, \Psi_{1,\alpha\beta} - \Psi_{2,\alpha\beta})$$
$$= \frac{1}{2}\begin{bmatrix} \Psi_{1\alpha} & \Psi_{1\beta} & \Psi_{2\alpha} & \Psi_{2\beta} \end{bmatrix} \cdot$$
$$\begin{bmatrix} L^{-1} & 0 & -L^{-1} & 0 \\ 0 & L^{-1} & 0 & -L^{-1} \\ -L^{-1} & 0 & L^{-1} & 0 \\ 0 & -L^{-1} & 0 & L^{-1} \end{bmatrix}\begin{bmatrix} \Psi_{1\alpha} \\ \Psi_{1\beta} \\ \Psi_{2\alpha} \\ \Psi_{2\beta} \end{bmatrix} \tag{7}$$

Let $\Psi = [\Psi_{1\alpha}, \Psi_{1\beta}, \Psi_{2\alpha}, \Psi_{2\beta}, \Psi_{3\alpha}, \Psi_{3\beta}, \Psi_f, \Psi_D, \Psi_g, \Psi_Q]^T$ be the state vector, (7) can be expressed as

$$E_l = \frac{1}{2}\Psi^T K_L \Psi \tag{8}$$

where $K_L = B^T \begin{bmatrix} L^{-1} & 0 & -L^{-1} & 0 \\ 0 & L^{-1} & 0 & -L^{-1} \\ -L^{-1} & 0 & L^{-1} & 0 \\ 0 & -L^{-1} & 0 & L^{-1} \end{bmatrix} B$ is the extended

matrix of the coefficient matrix in (7), and $B$ is the association matrix.

The relationship between winding currents and flux linkages of the synchronous generator shown in Fig.3b and Fig.3c in dq reference frame can be expressed by [7]:

$$\begin{bmatrix} \Psi_d \\ \Psi_q \\ \Psi_0 \\ \Psi_f \\ \Psi_D \\ \Psi_g \\ \Psi_Q \end{bmatrix} = \begin{bmatrix} L_d & 0 & 0 & L_{df} & L_{dD} & 0 & 0 \\ 0 & L_q & 0 & 0 & 0 & L_{qg} & L_{qQ} \\ 0 & 0 & L_0 & 0 & 0 & 0 & 0 \\ L_{fd} & 0 & 0 & L_f & L_{fD} & 0 & 0 \\ L_{Dd} & 0 & 0 & L_{fD} & L_D & 0 & 0 \\ 0 & L_{gq} & 0 & 0 & 0 & L_g & L_{gQ} \\ 0 & L_{Qq} & 0 & 0 & 0 & L_{gQ} & L_Q \end{bmatrix}\begin{bmatrix} i_d \\ i_q \\ i_0 \\ i_f \\ i_D \\ i_g \\ i_Q \end{bmatrix} \tag{9}$$

(9) is abbreviated as

$$\Psi_{dq0fDgQ} = L_{dq0}I_{dq0fDgQ} \tag{10}$$

By inverting the coefficient matrix in (9) and ignoring the zero-axis component, (9) is rewritten as

$$\begin{bmatrix} i_d \\ i_q \\ i_f \\ i_D \\ i_g \\ i_Q \end{bmatrix} = \begin{bmatrix} \Gamma_d & 0 & \Gamma_{df} & \Gamma_{dD} & 0 & 0 \\ 0 & \Gamma_q & 0 & 0 & \Gamma_{qg} & \Gamma_{qQ} \\ \Gamma_{fd} & 0 & \Gamma_f & \Gamma_{fD} & 0 & 0 \\ \Gamma_{Dd} & 0 & \Gamma_{fD} & \Gamma_D & 0 & 0 \\ 0 & \Gamma_{gq} & 0 & 0 & \Gamma_g & \Gamma_{gQ} \\ 0 & \Gamma_{Qq} & 0 & 0 & \Gamma_{gQ} & \Gamma_Q \end{bmatrix}\begin{bmatrix} \Psi_d \\ \Psi_q \\ \Psi_f \\ \Psi_D \\ \Psi_g \\ \Psi_Q \end{bmatrix} \tag{11}$$

(11) is abbreviated as

$$I_{dqfDgQ} = \Gamma_{dq}\Psi_{dqfDgQ} \tag{12}$$

The magnetic field energy stored in the generator can be expressed in the quadratic form of (13).

$$E_\mathrm{m} = \frac{1}{2}\boldsymbol{\Psi}_\mathrm{dqfDgQ}^\mathrm{T}\boldsymbol{\Gamma}_\mathrm{dq}\boldsymbol{\Psi}_\mathrm{dqfDgQ}$$
$$= \frac{1}{2}\boldsymbol{\Psi}_\mathrm{\alpha\beta fDgQ}^\mathrm{T}\left[\boldsymbol{P}(\boldsymbol{\theta})\boldsymbol{\Gamma}_\mathrm{dq}\boldsymbol{P}^{-1}(\boldsymbol{\theta})\right]\boldsymbol{\Psi}_\mathrm{\alpha\beta fDgQ} \quad (13)$$
$$= \frac{1}{2}\boldsymbol{\Psi}^\mathrm{T}\boldsymbol{\Gamma}(\boldsymbol{\theta})\boldsymbol{\Psi}$$

where 
$$\boldsymbol{P}(\boldsymbol{\theta}) = \begin{bmatrix} \cos\theta_5 & -\sin\theta_5 & 0 & 0 & 0 & 0 \\ \sin\theta_5 & \cos\theta_5 & 0 & 0 & 0 & 0 \\ 0 & 0 & 1 & 0 & 0 & 0 \\ 0 & 0 & 0 & 1 & 0 & 0 \\ 0 & 0 & 0 & 0 & 1 & 0 \\ 0 & 0 & 0 & 0 & 0 & 1 \end{bmatrix},$$

$\boldsymbol{\Gamma}(\boldsymbol{\theta}) = \boldsymbol{B}^\mathrm{T}\boldsymbol{P}(\boldsymbol{\theta})\boldsymbol{\Gamma}_\mathrm{dq}\boldsymbol{P}^{-1}(\boldsymbol{\theta})\boldsymbol{B}$. The energy $E_\mathrm{m}$ stored in the generator windings is not only related to the flux linkage vector, but also related to the angular displacement $\theta_5$ of the rotor of the generator. In general, the inverse inductance matrix $\boldsymbol{\Gamma}(\boldsymbol{\theta})$ is a matrix function of $\boldsymbol{\theta}$.

The potential energy of the rotor is
$$E_\mathrm{k} = \frac{1}{2}\boldsymbol{\theta}^\mathrm{T}\boldsymbol{K}\boldsymbol{\theta} \quad (14)$$
where $\boldsymbol{K}$ is the stiffness matrix of the mechanical shafting.

The total potential energy including the energy stored in the inductance branches in the circuit, the magnetic field energy of the generator, and the elastic potential energy of the mechanical shafting, i.e.,
$$U = E_l + E_\mathrm{m} + E_\mathrm{k}$$
$$= \frac{1}{2}\boldsymbol{\Psi}^\mathrm{T}\boldsymbol{K}_L\boldsymbol{\Psi} + \frac{1}{2}\boldsymbol{\Psi}^\mathrm{T}\boldsymbol{\Gamma}(\boldsymbol{\theta})\boldsymbol{\Psi} + \frac{1}{2}\boldsymbol{\theta}^\mathrm{T}\boldsymbol{K}\boldsymbol{\theta} \quad (15)$$

### B. Definition of Kinetic Energy

The kinetic energy of the system includes the energy stored in the capacitor branch and the rotational kinetic energy of the mechanical shafting. According to the same matrix notation as in section III.A, the capacitor energy can be written as
$$E_\mathrm{c} = \frac{1}{2}\dot{\boldsymbol{\Psi}}^\mathrm{T}\boldsymbol{K}_\mathrm{C}\dot{\boldsymbol{\Psi}} \quad (16)$$
where $\boldsymbol{K}_\mathrm{C}$ is the capacitance coefficient matrix. The rotational kinetic energy of the mechanical shafting is
$$E_\mathrm{J} = \frac{1}{2}\dot{\boldsymbol{\theta}}^\mathrm{T}\boldsymbol{J}\dot{\boldsymbol{\theta}} \quad (17)$$
where $\boldsymbol{J}$ is the moment of inertia matrix of shaft system. The total kinetic energy of the system is
$$T = E_\mathrm{c} + E_\mathrm{J}$$
$$= \frac{1}{2}\dot{\boldsymbol{\Psi}}^\mathrm{T}\boldsymbol{K}_\mathrm{C}\dot{\boldsymbol{\Psi}} + \frac{1}{2}\dot{\boldsymbol{\theta}}^\mathrm{T}\boldsymbol{J}\dot{\boldsymbol{\theta}} \quad (18)$$

### C. Lagrangian Function and Lagrangian ODE Modelling

The Lagrangian function is defined as kinetic energy minus potential energy of the system, i.e., $\mathcal{L} = T - U$. Therefore, the Lagrangian function of the power system shown in Fig.3 is written as

$$\mathcal{L}(\dot{\boldsymbol{\Psi}},\dot{\boldsymbol{\theta}},\boldsymbol{\Psi},\boldsymbol{\theta}) = \left(\frac{1}{2}\dot{\boldsymbol{\Psi}}^\mathrm{T}\boldsymbol{K}_C\dot{\boldsymbol{\Psi}} + \frac{1}{2}\dot{\boldsymbol{\theta}}^\mathrm{T}\boldsymbol{J}\dot{\boldsymbol{\theta}}\right)$$
$$-\left(\frac{1}{2}\boldsymbol{\Psi}^\mathrm{T}\boldsymbol{K}_L\boldsymbol{\Psi} + \frac{1}{2}\boldsymbol{\Psi}^\mathrm{T}\boldsymbol{\Gamma}(\boldsymbol{\theta})\boldsymbol{\Psi} + \frac{1}{2}\boldsymbol{\theta}^\mathrm{T}\boldsymbol{K}\boldsymbol{\theta}\right) \quad (19)$$

Defining the generalized coordinate of the system as $\boldsymbol{x} = \begin{bmatrix}\boldsymbol{\Psi}\\\boldsymbol{\theta}\end{bmatrix}$, (19) is transformed into the following form.

$$\mathcal{L}(\dot{\boldsymbol{x}},\boldsymbol{x}) = \frac{1}{2}\dot{\boldsymbol{x}}^\mathrm{T}\begin{bmatrix}\boldsymbol{K}_C & 0\\ 0 & \boldsymbol{J}\end{bmatrix}\dot{\boldsymbol{x}} - \frac{1}{2}\boldsymbol{x}^\mathrm{T}\begin{bmatrix}\boldsymbol{K}_L+\boldsymbol{\Gamma}(\boldsymbol{\theta}) & 0\\ 0 & \boldsymbol{K}\end{bmatrix}\boldsymbol{x} \quad (20)$$

Substituting (20) into the Eular-Lagrangian equation shown in (21), the ODE given by (22) is obtained.
$$\frac{d}{dt}\left(\frac{\partial\mathcal{L}}{\partial\dot{\boldsymbol{x}}}\right) - \frac{\partial\mathcal{L}}{\partial\boldsymbol{x}} = 0 \quad (21)$$

$$\begin{bmatrix}\boldsymbol{K}_C & 0\\ 0 & \boldsymbol{J}\end{bmatrix}\begin{bmatrix}\ddot{\boldsymbol{\Psi}}\\ \ddot{\boldsymbol{\theta}}\end{bmatrix} + \begin{bmatrix}\boldsymbol{K}_L & 0\\ 0 & \boldsymbol{K}\end{bmatrix}\begin{bmatrix}\boldsymbol{\Psi}\\ \boldsymbol{\theta}\end{bmatrix} + \begin{bmatrix}\boldsymbol{\Gamma}(\boldsymbol{\theta})\boldsymbol{\Psi}\\ \frac{1}{2}\boldsymbol{\Psi}^\mathrm{T}\boldsymbol{\Gamma}'(\boldsymbol{\theta})\boldsymbol{\Psi}\end{bmatrix} = 0 \quad (22)$$

where $\boldsymbol{\Gamma}'(\boldsymbol{\theta}) = \frac{\partial\boldsymbol{\Gamma}(\boldsymbol{\theta})}{\partial\boldsymbol{\theta}}$, expansion of which is shown in Appendix A.

Considering the dissipation of resistance and friction damping, as well as the effect of external torque and ideal power supply, the complete ODE model of the system is

$$\begin{bmatrix}\boldsymbol{K}_C & 0\\ 0 & \boldsymbol{J}\end{bmatrix}\begin{bmatrix}\ddot{\boldsymbol{\Psi}}\\ \ddot{\boldsymbol{\theta}}\end{bmatrix} + \begin{bmatrix}\boldsymbol{K}_R & 0\\ 0 & \boldsymbol{D}\end{bmatrix}\begin{bmatrix}\dot{\boldsymbol{\Psi}}\\ \dot{\boldsymbol{\theta}}\end{bmatrix} +$$
$$\begin{bmatrix}\boldsymbol{K}_L & 0\\ 0 & \boldsymbol{K}\end{bmatrix}\begin{bmatrix}\boldsymbol{\Psi}\\ \boldsymbol{\theta}\end{bmatrix} + \begin{bmatrix}\boldsymbol{\Gamma}(\boldsymbol{\theta})\boldsymbol{\Psi}\\ \frac{1}{2}\boldsymbol{\Psi}^\mathrm{T}\boldsymbol{\Gamma}'(\boldsymbol{\theta})\boldsymbol{\Psi}\end{bmatrix} = \begin{bmatrix}\boldsymbol{i}_\mathrm{s}\\ \boldsymbol{T}\end{bmatrix} \quad (23)$$

where $\boldsymbol{i}_\mathrm{s}$ represents the vector of current injection of nodes, $\boldsymbol{i}_s = \left[\frac{u_\mathrm{s}}{R}\cos(\omega_\mathrm{s}t), \frac{u_\mathrm{s}}{R}\sin(\omega_\mathrm{s}t), 0,0,0,0, \frac{u_\mathrm{f}}{R_\mathrm{f}}, 0,0,0\right]^\mathrm{T}$, where the nonzero quantities represent the injection currents of the infinite power source and the excitation power source; $\boldsymbol{T} = [T_1, T_2, T_3, T_4, 0, 0]^\mathrm{T}$ is the vector of mechanical torque on the shaft system; the electromagnetic torque on the shaft is $T_\mathrm{E} = -\frac{1}{2}\boldsymbol{\Psi}^\mathrm{T}\boldsymbol{\Gamma}'(\boldsymbol{\theta})\boldsymbol{\Psi}$; $\boldsymbol{K}_R$ is the resistance coefficient matrix; and $\boldsymbol{D}$ is the damping coefficient matrix of the mechanical shafting.

## IV. TRANSFORM TO ZERO SOLUTION STABILITY PROBLEM

### A. Transform to Synchronous Coordinate System

Equation (23) is a second-order matrix ODE, which could be transformed into a standard form like (1) by variable substitution $\boldsymbol{x} = \begin{bmatrix}\dot{\boldsymbol{\Psi}}\\ \dot{\boldsymbol{\theta}}\\ \boldsymbol{\Psi}\\ \boldsymbol{\theta}\end{bmatrix}$. The node flux $\Psi_{1\alpha}$, $\Psi_{1\beta}$, $\Psi_{2\alpha}$, $\Psi_{2\beta}$, $\Psi_{3\alpha}$, $\Psi_{3\beta}$ and node voltage $\dot{\Psi}_{1\alpha}$, $\dot{\Psi}_{1\beta}$, $\dot{\Psi}_{2\alpha}$, $\dot{\Psi}_{2\beta}$, $\dot{\Psi}_{3\alpha}$, $\dot{\Psi}_{3\beta}$ in $x(t)$ are AC sinusoidal quantities. Although it could be analyzed by Lyapunov's first method directly, it is a common practice to transform (23) from the stationary αβ coordinate system to the synchronously rotating coordinate system in power system analysis. On the synchronously rotating coordinate system, the problem is an equilibrium stability problem.

Let the x-axis of the synchronous reference frame be aligned to the voltage vector of the infinite bus in Fig.3, the xy synchronous coordinate system is established. Let the generalized coordinate vector in the xy synchronous coordinate system be $\boldsymbol{\varphi}=(\varphi_{1x}, \varphi_{1y}, \varphi_{2x}, \varphi_{2y}, \varphi_{3x}, \varphi_{3y}, \Psi_f, \Psi_D, \Psi_g, \Psi_Q)^T$, the generalized coordinate vector of mechanical shaft be $\boldsymbol{\delta}=(\delta_1, \delta_2, \delta_3, \delta_4, \delta_5, \delta_6)$. The flux linkages $\Psi_f, \Psi_D, \Psi_g, \Psi_Q$ of the rotor windings of the generator are constant at the equilibrium operating point. The relationship between the node flux linkage $[\Psi_\alpha, \Psi_\beta]^T$ in the stationary αβ coordinate system and that $[\varphi_x, \varphi_y]^T$ in the xy synchronously rotating coordinate system can be expressed as follows.

$$\begin{bmatrix} \Psi_\alpha \\ \Psi_\beta \end{bmatrix} = \boldsymbol{K}_{rot}(\theta_s) \begin{bmatrix} \varphi_x \\ \varphi_y \end{bmatrix} \quad (24)$$

where $\boldsymbol{K}_{rot}(\theta_s) = \begin{bmatrix} \cos(\theta_s) & -\sin(\theta_s) \\ \sin(\theta_s) & \cos(\theta_s) \end{bmatrix}$, and $\theta_s$ is the phase angle of the voltage vector of the infinite bus.

Consider $\frac{d}{d\theta_s}\boldsymbol{K}_{rot}(\theta_s) = \boldsymbol{K}_j \boldsymbol{K}_{rot}(\theta_s)$, and $\boldsymbol{K}_j = \begin{bmatrix} 0 & -1 \\ 1 & 0 \end{bmatrix}$, we have (25) and (26).

$$\begin{bmatrix} \dot{\Psi}_\alpha \\ \dot{\Psi}_\beta \end{bmatrix} = \boldsymbol{K}_{rot}(\theta_s) \begin{bmatrix} \dot{\varphi}_x \\ \dot{\varphi}_y \end{bmatrix} + \dot{\theta}_s \boldsymbol{K}_j \boldsymbol{K}_{rot}(\theta_s) \begin{bmatrix} \varphi_x \\ \varphi_y \end{bmatrix} \quad (25)$$

$$\begin{bmatrix} \boldsymbol{K}_C & 0 \\ 0 & \boldsymbol{J} \end{bmatrix}\begin{bmatrix} \ddot{\boldsymbol{\varphi}} \\ \ddot{\boldsymbol{\delta}} \end{bmatrix} + \begin{bmatrix} \boldsymbol{K}_R + 2\omega_s\boldsymbol{K}_j\boldsymbol{K}_C & 0 \\ 0 & \boldsymbol{D} \end{bmatrix}\begin{bmatrix} \dot{\boldsymbol{\varphi}} \\ \dot{\boldsymbol{\delta}} \end{bmatrix} + \begin{bmatrix} \boldsymbol{K}_L + \omega_s\boldsymbol{K}_j\boldsymbol{K}_R - \omega_s^2\boldsymbol{K}_C & 0 \\ 0 & \boldsymbol{K} \end{bmatrix}\begin{bmatrix} \boldsymbol{\varphi} \\ \boldsymbol{\delta} \end{bmatrix}$$

From (30), the standard form of (1) could be obtained by using variable substitution $\boldsymbol{x} = \begin{bmatrix} \dot{\boldsymbol{\varphi}} \\ \dot{\boldsymbol{\delta}} \\ \boldsymbol{\varphi} \\ \boldsymbol{\delta} \end{bmatrix}$, and the motion stability problem could be transformed into the equilibrium stability problem $\boldsymbol{x}_0 = \begin{bmatrix} \dot{\boldsymbol{\varphi}}_0 \\ \dot{\boldsymbol{\delta}}_0 \\ \boldsymbol{\varphi}_0 \\ \boldsymbol{\delta}_0 \end{bmatrix}$. The acceleration vector of the stationary equilibrium operating point in the synchronous coordinate system must be zero vectors, i.e. $\ddot{\boldsymbol{\varphi}}_0 = 0$ and $\ddot{\boldsymbol{\delta}}_0 = 0$. The velocity vector must also be zero vectors, i.e. $\dot{\boldsymbol{\varphi}}_0 = 0$ and $\dot{\boldsymbol{\delta}}_0 = 0$. Therefore, we have (31).

$$\begin{bmatrix} \ddot{\Psi}_\alpha \\ \ddot{\Psi}_\beta \end{bmatrix} = \boldsymbol{K}_{rot}(\theta_s) \begin{bmatrix} \ddot{\varphi}_x \\ \ddot{\varphi}_y \end{bmatrix} + 2\dot{\theta}_s \boldsymbol{K}_j \boldsymbol{K}_{rot}(\theta_s) \begin{bmatrix} \dot{\varphi}_x \\ \dot{\varphi}_y \end{bmatrix} + \dot{\theta}_s^2 \boldsymbol{K}_j^2 \boldsymbol{K}_{rot}(\theta_s) \begin{bmatrix} \varphi_x \\ \varphi_y \end{bmatrix} \quad (26)$$

The relative angular displacement vector $\boldsymbol{\delta}$ between the rotational masses and the phase angle $\theta_s$ of the infinite voltage source is

$$\begin{cases} \boldsymbol{\theta} = \boldsymbol{\delta} + \boldsymbol{\theta}_s \\ \dot{\boldsymbol{\theta}} = \dot{\boldsymbol{\delta}} + \boldsymbol{\omega}_s \\ \ddot{\boldsymbol{\theta}} = \ddot{\boldsymbol{\delta}} \end{cases} \quad (27)$$

where $\boldsymbol{\delta} = \begin{bmatrix} \theta_1 - \theta_s \\ \theta_2 - \theta_s \\ \theta_3 - \theta_s \\ \theta_4 - \theta_s \\ \theta_5 - \theta_s \\ \theta_6 - \theta_s \end{bmatrix}$, $\dot{\boldsymbol{\theta}}_s = \boldsymbol{\omega}_s = \begin{bmatrix} \omega_s \\ \omega_s \\ \omega_s \\ \omega_s \\ \omega_s \\ \omega_s \end{bmatrix}$.

Substituting (24)-(27) into (23), we have the ODEs given by (30) and (31) in xy synchronously rotating coordinate system.

$$\boldsymbol{K}_{rot}(\theta_s)\{\boldsymbol{K}_C\ddot{\boldsymbol{\varphi}} + [\boldsymbol{K}_R + 2\omega_s\boldsymbol{K}_j\boldsymbol{K}_C]\dot{\boldsymbol{\varphi}} + [\boldsymbol{K}_L + \omega_s\boldsymbol{K}_j\boldsymbol{K}_R - \omega_s^2\boldsymbol{K}_C]\boldsymbol{\varphi} + \boldsymbol{\Gamma}(\boldsymbol{\delta})\boldsymbol{\varphi}\} = \boldsymbol{K}_{rot}(\theta_s)\boldsymbol{I}_{ss} \quad (28)$$

$$\boldsymbol{J}\ddot{\boldsymbol{\delta}} + \boldsymbol{D}\dot{\boldsymbol{\delta}} + \boldsymbol{K}\boldsymbol{\delta} + \left[\frac{1}{2}\boldsymbol{\varphi}^T \frac{\partial \boldsymbol{\Gamma}(\boldsymbol{\delta})}{\partial \boldsymbol{\delta}}\boldsymbol{\varphi} + \boldsymbol{D}\boldsymbol{\omega}_s\right] = \boldsymbol{T} \quad (29)$$

where $\boldsymbol{I}_{ss} = \left[\frac{u_s}{R}, 0, 0, 0, 0, 0, \frac{u_f}{R_f}, 0, 0, 0\right]^T$. Matrix $\boldsymbol{K}_{rot}(\theta_s)$ is eliminated on both sides of (28). Combined with (29), we have the ODEs given in (30).

$$+ \begin{bmatrix} \boldsymbol{\Gamma}(\boldsymbol{\delta})\boldsymbol{\varphi} \\ \frac{1}{2}\boldsymbol{\varphi}^T \boldsymbol{\Gamma}'(\boldsymbol{\delta})\boldsymbol{\varphi} + \boldsymbol{D}\boldsymbol{\omega}_s \end{bmatrix} = \begin{bmatrix} \boldsymbol{I}_{ss} \\ \boldsymbol{T} \end{bmatrix} \quad (30)$$

$$\begin{cases} (\boldsymbol{K}_L + \omega_s\boldsymbol{K}_j\boldsymbol{K}_R - \omega_s^2\boldsymbol{K}_C + \boldsymbol{\Gamma}(\boldsymbol{\delta}_0))\boldsymbol{\varphi}_0 = \boldsymbol{I}_{ss} \\ \boldsymbol{K}\boldsymbol{\delta}_0 + \frac{1}{2}\boldsymbol{\varphi}_0^T \boldsymbol{\Gamma}'(\boldsymbol{\delta}_0)\boldsymbol{\varphi}_0 + \boldsymbol{D}\boldsymbol{\omega}_s = \boldsymbol{T} \end{cases} \quad (31)$$

By solving the nonlinear algebraic equations (31), the stationary equilibrium operating point $\boldsymbol{\varphi}_0$ and $\boldsymbol{\delta}_0$ could be obtained.

### B. Transform to Zero Solution Stability Problem

With $\boldsymbol{\varphi}_0$ and $\boldsymbol{\delta}_0$, the equilibrium stability problem could be transformed into the origin stability problem. In the synchronously rotating coordinate system, $\boldsymbol{z}(t)$ is defined as the motion corresponding to the initial condition $\boldsymbol{x}_0 + \Delta\boldsymbol{x}_0$, i.e.,

$$z(t) = \begin{bmatrix} \dot{\boldsymbol{\varphi}}(t) \\ \dot{\boldsymbol{\delta}}(t) \\ \boldsymbol{\varphi}(t) \\ \boldsymbol{\delta}(t) \end{bmatrix} \text{ and } \begin{cases} \boldsymbol{\varphi}(t) = \boldsymbol{\varphi}_0 + \Delta\boldsymbol{\varphi}(t) \\ \boldsymbol{\delta}(t) = \boldsymbol{\delta}_0 + \Delta\boldsymbol{\delta}(t) \\ \dot{\boldsymbol{\varphi}}(t) = \Delta\dot{\boldsymbol{\varphi}}(t) \\ \dot{\boldsymbol{\delta}}(t) = \Delta\dot{\boldsymbol{\delta}}(t) \\ \ddot{\boldsymbol{\varphi}}(t) = \Delta\ddot{\boldsymbol{\varphi}}(t) \\ \ddot{\boldsymbol{\delta}}(t) = \Delta\ddot{\boldsymbol{\delta}}(t) \end{cases} \quad (32)$$

According to the definition of motion stability offered in Section II, it is necessary to obtain the differential equation with $\Delta x = z - x_0$ as the state variable. Substitute (32) into (30), we have the small-signal model with $\Delta x = \begin{bmatrix} \Delta\dot{\boldsymbol{\varphi}} \\ \Delta\dot{\boldsymbol{\delta}} \\ \Delta\boldsymbol{\varphi} \\ \Delta\boldsymbol{\delta} \end{bmatrix}$ being the state variable, as shown in (33). In (33), $\boldsymbol{\varphi}_0$ and $\boldsymbol{\delta}_0$ are known constants of the equilibrium state. Using (31) to eliminate $\boldsymbol{I}_{ss}$, we get (34).

$$\begin{bmatrix} \boldsymbol{K}_C & 0 \\ 0 & \boldsymbol{J} \end{bmatrix}\begin{bmatrix} \Delta\ddot{\boldsymbol{\varphi}} \\ \Delta\ddot{\boldsymbol{\delta}} \end{bmatrix} + \begin{bmatrix} \boldsymbol{K}_R + 2\omega_s \boldsymbol{K}_j \boldsymbol{K}_C & 0 \\ 0 & \boldsymbol{D} \end{bmatrix}\begin{bmatrix} \Delta\dot{\boldsymbol{\varphi}} \\ \Delta\dot{\boldsymbol{\delta}} \end{bmatrix} + \begin{bmatrix} \boldsymbol{K}_L + \omega_s \boldsymbol{K}_j \boldsymbol{K}_R - \omega_s^2 \boldsymbol{K}_C & 0 \\ 0 & \boldsymbol{K} \end{bmatrix}\begin{bmatrix} \boldsymbol{\varphi}_0 + \Delta\boldsymbol{\varphi}(t) \\ \boldsymbol{\delta}_0 + \Delta\boldsymbol{\delta} \end{bmatrix}$$
$$+ \begin{bmatrix} \boldsymbol{\Gamma}(\boldsymbol{\delta}_0 + \Delta\boldsymbol{\delta}(t))(\boldsymbol{\varphi}_0 + \Delta\boldsymbol{\varphi}(t)) \\ \frac{1}{2}(\boldsymbol{\varphi}_0 + \Delta\boldsymbol{\varphi}(t))^T \boldsymbol{\Gamma}'(\boldsymbol{\delta}_0 + \Delta\boldsymbol{\delta}(t))(\boldsymbol{\varphi}_0 + \Delta\boldsymbol{\varphi}(t)) + \boldsymbol{D}\omega_s \end{bmatrix} = \begin{bmatrix} \boldsymbol{I}_{ss} \\ \boldsymbol{T} \end{bmatrix} \quad (33)$$

$$\begin{bmatrix} \boldsymbol{K}_C & 0 \\ 0 & \boldsymbol{J} \end{bmatrix}\begin{bmatrix} \Delta\ddot{\boldsymbol{\varphi}} \\ \Delta\ddot{\boldsymbol{\delta}} \end{bmatrix} + \begin{bmatrix} \boldsymbol{K}_R + 2\omega_s \boldsymbol{K}_j \boldsymbol{K}_C & 0 \\ 0 & \boldsymbol{D} \end{bmatrix}\begin{bmatrix} \Delta\dot{\boldsymbol{\varphi}} \\ \Delta\dot{\boldsymbol{\delta}} \end{bmatrix} + \begin{bmatrix} \boldsymbol{K}_L + \omega_s \boldsymbol{K}_j \boldsymbol{K}_R - \omega_s^2 \boldsymbol{K}_C & 0 \\ 0 & \boldsymbol{K} \end{bmatrix}\begin{bmatrix} \Delta\boldsymbol{\varphi} \\ \Delta\boldsymbol{\delta} \end{bmatrix}$$
$$= \begin{bmatrix} \boldsymbol{\Gamma}(\boldsymbol{\delta}_0)\boldsymbol{\varphi}_0(t) - \boldsymbol{\Gamma}(\boldsymbol{\delta}_0 + \Delta\boldsymbol{\delta}(t))(\boldsymbol{\varphi}_0 + \Delta\boldsymbol{\varphi}(t)) \\ \frac{1}{2}\boldsymbol{\varphi}_0^T \boldsymbol{\Gamma}'(\boldsymbol{\delta}_0)\boldsymbol{\varphi}_0 - \frac{1}{2}(\boldsymbol{\varphi}_0 + \Delta\boldsymbol{\varphi}(t))^T \boldsymbol{\Gamma}'(\boldsymbol{\delta}_0 + \Delta\boldsymbol{\delta}(t))(\boldsymbol{\varphi}_0 + \Delta\boldsymbol{\varphi}(t)) \end{bmatrix} \quad (34)$$

$$\begin{bmatrix} \boldsymbol{K}_C & 0 & 0 & 0 \\ 0 & \boldsymbol{J} & 0 & 0 \\ 0 & 0 & \boldsymbol{E} & 0 \\ 0 & 0 & 0 & \boldsymbol{E} \end{bmatrix}\begin{bmatrix} \Delta\ddot{\boldsymbol{\varphi}} \\ \Delta\ddot{\boldsymbol{\delta}} \\ \Delta\dot{\boldsymbol{\varphi}} \\ \Delta\dot{\boldsymbol{\delta}} \end{bmatrix} = -\begin{bmatrix} \boldsymbol{K}_R + 2\omega_s \boldsymbol{K}_j \boldsymbol{K}_C & 0 & \boldsymbol{K}_L + \omega_s \boldsymbol{K}_j \boldsymbol{K}_R - \omega_s^2 \boldsymbol{K}_C & 0 \\ 0 & \boldsymbol{D} & 0 & \boldsymbol{K} \\ -\boldsymbol{E} & 0 & 0 & 0 \\ 0 & -\boldsymbol{E} & 0 & 0 \end{bmatrix}\begin{bmatrix} \Delta\dot{\boldsymbol{\varphi}} \\ \Delta\dot{\boldsymbol{\delta}} \\ \Delta\boldsymbol{\varphi} \\ \Delta\boldsymbol{\delta} \end{bmatrix}$$
$$+ \begin{bmatrix} \boldsymbol{\Gamma}(\boldsymbol{\delta}_0)\boldsymbol{\varphi}_0 - \boldsymbol{\Gamma}(\boldsymbol{\delta}_0 + \Delta\boldsymbol{\delta}(t))(\boldsymbol{\varphi}_0 + \Delta\boldsymbol{\varphi}(t)) \\ \frac{1}{2}\boldsymbol{\varphi}_0^T \boldsymbol{\Gamma}'(\boldsymbol{\delta}_0)\boldsymbol{\varphi}_0 - \frac{1}{2}(\boldsymbol{\varphi}_0 + \Delta\boldsymbol{\varphi}(t))^T \boldsymbol{\Gamma}'(\boldsymbol{\delta}_0 + \Delta\boldsymbol{\delta}(t))(\boldsymbol{\varphi}_0 + \Delta\boldsymbol{\varphi}(t)) \\ 0 \\ 0 \end{bmatrix} \quad (35)$$

$$\boldsymbol{\Gamma}(\boldsymbol{\delta}_0 + \Delta\boldsymbol{\delta}(t))(\boldsymbol{\varphi}_0 + \Delta\boldsymbol{\varphi}(t)) = \boldsymbol{\Gamma}(\boldsymbol{\delta}_0)\boldsymbol{\varphi}_0 + \Delta\boldsymbol{\delta}(t)\boldsymbol{\Gamma}'(\boldsymbol{\delta}_0)\boldsymbol{\varphi}_0 + \boldsymbol{\Gamma}(\boldsymbol{\delta}_0)\Delta\boldsymbol{\varphi}(t) + \cdots \quad (36)$$

$$\frac{1}{2}(\boldsymbol{\varphi}_0 + \Delta\boldsymbol{\varphi}(t))^T \boldsymbol{\Gamma}'(\boldsymbol{\delta}_0 + \Delta\boldsymbol{\delta}(t))(\boldsymbol{\varphi}_0 + \Delta\boldsymbol{\varphi}(t)) = \frac{1}{2}\boldsymbol{\varphi}_0^T \boldsymbol{\Gamma}'(\boldsymbol{\delta}_0)\boldsymbol{\varphi}_0 + \frac{1}{2}\boldsymbol{\varphi}_0^T \boldsymbol{\Gamma}''(\boldsymbol{\delta}_0)\boldsymbol{\varphi}_0 \Delta\boldsymbol{\delta}(t) + \boldsymbol{\varphi}_0^T \boldsymbol{\Gamma}'(\boldsymbol{\delta}_0)\Delta\boldsymbol{\varphi}(t) + \cdots \quad (37)$$

where $\boldsymbol{\Gamma}''(\boldsymbol{\delta}_0) = \dfrac{\partial^2 \boldsymbol{\Gamma}(\boldsymbol{\delta}_0)}{\partial \boldsymbol{\delta}_0^2}$, its expansion is shown in Appendix A.

$$\begin{bmatrix} \Delta\ddot{\boldsymbol{\varphi}} \\ \Delta\ddot{\boldsymbol{\delta}} \\ \Delta\dot{\boldsymbol{\varphi}} \\ \Delta\dot{\boldsymbol{\delta}} \end{bmatrix} = \left\{ -\begin{bmatrix} \boldsymbol{K}_C & 0 & 0 & 0 \\ 0 & \boldsymbol{J} & 0 & 0 \\ 0 & 0 & \boldsymbol{E} & 0 \\ 0 & 0 & 0 & \boldsymbol{E} \end{bmatrix}^{-1} \begin{bmatrix} \boldsymbol{K}_R + 2\omega_s \boldsymbol{K}_j \boldsymbol{K}_C & 0 & \boldsymbol{K}_L + \omega_s \boldsymbol{K}_j \boldsymbol{K}_R - \omega_s^2 \boldsymbol{K}_C + \boldsymbol{\Gamma}(\boldsymbol{\delta}_0) & \boldsymbol{\Gamma}'(\boldsymbol{\delta}_0)\boldsymbol{\varphi}_0 \\ 0 & \boldsymbol{D} & \boldsymbol{\varphi}_0^T \boldsymbol{\Gamma}'(\boldsymbol{\delta}_0) & \boldsymbol{K} + \frac{1}{2}\boldsymbol{\varphi}_0^T \boldsymbol{\Gamma}''(\boldsymbol{\delta}_0)\boldsymbol{\varphi}_0 \\ -\boldsymbol{E} & 0 & 0 & 0 \\ 0 & -\boldsymbol{E} & 0 & 0 \end{bmatrix} \right\} \begin{bmatrix} \Delta\dot{\boldsymbol{\varphi}} \\ \Delta\dot{\boldsymbol{\delta}} \\ \Delta\boldsymbol{\varphi} \\ \Delta\boldsymbol{\delta} \end{bmatrix} \quad (38)$$

*C. Modal Analysis*

Transform (34) to (35), and it meets the form $d\Delta x/dt = g(\Delta x(t), t)$ required by step 2 in Section II-A. The coefficient matrix of (35) is function of $\boldsymbol{\varphi}_0$ and $\boldsymbol{\delta}_0$, it needs to be locally linearized. Ignoring the second-order and above infinitesimal terms on the right side of (36) and (37), substitute (36) and (37) into (35), then (38) is obtained. The eigenvalue analysis of the coefficient matrix in (38) could be used to judge the system's stability at the equilibrium operating point.

The output voltage of the generator is 15.01 kV. The output current is 19.82 kA, and the power factor is 0.9. The obtained $\varphi_0$ and $\delta_0$ are shown in Table 1.

Table 1 Equilibrium point

| Electrical node | Node flux linkage $\varphi_0$, unit Wb | | Mass block number | Relative angular displacement $\delta_0$, unit rad |
|---|---|---|---|---|
| | x-axis component | y-axis component | | |
| 1 | -0.0064229 | -62.692 | 1 | -0.48405 |
| 2 | -26.577 | -58.906 | 2 | -0.49804 |
| 3 | 21.002 | -65.688 | 3 | -0.51247 |
| f winding | 467.94 | | 4 | -0.52596 |
| d winding | 424.22 | | 5 | -0.53866 |
| q winding | -298.00 | | 6 | -0.53866 |
| g winding | -298.00 | | | |

The eigenvalues of the system are shown in Table 2, and the modal information corresponding to the No. 11 eigenvalue is shown in Table 3.

Table 2 Eigenvalue and eigen-frequency

| Order number | Eigenvalue | Eigen-frequency (Hz) |
|---|---|---|
| 1,2 | -1.59E-07±298.18i | 47.46 |
| 3,4 | -0.0022±202.78i | 32.27 |
| 5,6 | -0.0076±160.36i | 25.52 |
| 7,8 | -0.31±126.46i | 20.13 |
| 9,10 | -0.141±100.32i | 15.97 |
| 11,12 | 2.15±123.42i | 19.64 |
| 13,14 | -0.81±10.37i | 1.65 |
| 15,16 | -0.52±630.16i | 100.29 |
| 17,18 | -2.00E13±376.99i | 60.00 |
| 19,20 | -8.50E-05±376.99i | 60.00 |
| 21,22 | -4.05±4.69E6i | 7.46E5 |
| 23,24 | -3.81±4.00E6i | 6.37E5 |
| 25 | -1.88E11+0.00i | 0 |
| 26 | -6.46E10+0.00i | 0 |
| 27 | -3.21E10+0.00i | 0 |
| 28 | -1.88E10+0.00i | 0 |
| 29 | -33.17+0.00i | 0 |
| 30 | -20.44+0.0i | 0 |
| 31 | -0.58+0.00i | 0 |
| 32 | -3.50+0.00i | 0 |

Note: in the above table, E represents the scientific counting method.

Table 3 Modal vectors of the No. 11 eigenvalue

| Order number | Symbols | Modal vector |
|---|---|---|
| 1 | $\Delta\dot{\varphi}_{1x}$ | -4.27E-05+0.000945i |
| 2 | $\Delta\dot{\varphi}_{1y}$ | -0.000967-7.70E-06i |
| 3 | $\Delta\dot{\varphi}_{2x}$ | 0.694+0.00i |
| 4 | $\Delta\dot{\varphi}_{2y}$ | -0.0492+0.663i |
| 5 | $\Delta\dot{\varphi}_{3x}$ | -0.149-0.00725i |
| 6 | $\Delta\dot{\varphi}_{3y}$ | -0.0264-0.170i |
| 7 | $\Delta\dot{\Psi}_{f}$ | 0.00592+0.00118i |
| 8 | $\Delta\dot{\Psi}_{D}$ | 0.0237+0.0759i |
| 9 | $\Delta\dot{\Psi}_{g}$ | -0.0519+0.0336i |
| 10 | $\Delta\dot{\Psi}_{Q}$ | -0.109+0.0523i |
| 11 | $\Delta\omega_1$ | -0.000191-0.000292i |
| 12 | $\Delta\omega_2$ | -0.000113-0.000181i |
| 13 | $\Delta\omega_3$ | -2.66E-05-5.60E-05i |
| 14 | $\Delta\omega_4$ | 6.92E-05+0.000101i |
| 15 | $\Delta\omega_5$ | 6.63E-05+0.000117i |
| 16 | $\Delta\omega_6$ | 0.00325+1.23E-05i |
| 17 | $\Delta\varphi_{1x}$ | 7.65E-06+4.79E-07i |
| 18 | $\Delta\varphi_{1y}$ | -1.99E-07+7.83E-06i |
| 19 | $\Delta\varphi_{2x}$ | 9.78E-05-0.00562i |
| 20 | $\Delta\varphi_{2y}$ | 0.00536+0.000492i |
| 21 | $\Delta\varphi_{3x}$ | -7.98E-05+0.00121i |
| 22 | $\Delta\varphi_{3y}$ | -0.00138+0.000192i |
| 23 | $\Delta\Psi_{f}$ | 1.038E-05-4.78E-05i |
| 24 | $\Delta\Psi_{D}$ | 0.000618-0.000181i |
| 25 | $\Delta\Psi_{g}$ | 0.000265+0.000425i |
| 26 | $\Delta\Psi_{Q}$ | 0.000409+0.000889i |
| 27 | $\Delta\theta_1$ | -2.39E-06+1.51E-06i |
| 28 | $\Delta\theta_2$ | -1.48E-06+8.87E-07i |
| 29 | $\Delta\theta_3$ | -4.58E-07+2.07E-07i |
| 30 | $\Delta\theta_4$ | 8.30E-07-5.47E-07i |
| 31 | $\Delta\theta_5$ | 9.58E-07-5.20E-07i |
| 32 | $\Delta\theta_6$ | 5.58E-07-2.63E-05i |

In Table 2, the five natural frequency points of the mechanical shafts (15.71Hz, 20.21Hz, 25.55Hz, 32.29Hz, 47.46Hz, respectively) have been shifted, which is caused by the coupling of the mechanical shafting with the generator and the electrical system. The No.1~ No.10 eigenvalues reflect the information of these five natural frequencies. The oscillation mode of the mechanical shaft system could also be examined by checking the modal vector. At these frequency points, the real part of the eigenvalue is still negative. The unstable frequency point is 19.64 Hz, as shown in Table 2, the 11th and 12th eigenvalue, which is a new oscillatory frequency point.

Table 2 shows the modal vectors corresponding to the 11th eigenvalue. Since the 11th and 12th eigenvalues are conjugates, only the modal vectors of 11th eigenvalue is given here. In the modal vectors shown in Table 2, every element is nonzero, which means that in the xy coordinate system, every state in the vector $\Delta \boldsymbol{x} = \begin{bmatrix} \Delta \dot{\boldsymbol{\varphi}} \\ \Delta \dot{\boldsymbol{\delta}} \\ \Delta \boldsymbol{\varphi} \\ \Delta \boldsymbol{\delta} \end{bmatrix}$ oscillates divergently at a frequency of 19.64Hz.

## V. NUMERICAL SOLUTIONS

Taking $\boldsymbol{x}_{\alpha\beta\_0}$ given by (39) as the initial point, the numerical solution for the system shown in Fig.3 can be obtained with (23).

$$\boldsymbol{x}_{\alpha\beta\_0} = \begin{bmatrix} \dot{\boldsymbol{\Psi}}_0 \\ \dot{\boldsymbol{\theta}}_0 \\ \boldsymbol{\Psi}_0 \\ \boldsymbol{\theta}_0 \end{bmatrix} = \begin{bmatrix} \omega_s \boldsymbol{K}_j \boldsymbol{K}_{rot}(\omega_s t) \cdot \boldsymbol{\varphi}_0 \\ \omega_s \\ \boldsymbol{K}_{rot}(\omega_s t) \cdot \boldsymbol{\varphi}_0 \\ \boldsymbol{\delta}_0 + \omega_s t \end{bmatrix}_{t=0} \quad (39)$$

The definition of matrix $\boldsymbol{K}_j$ and $\boldsymbol{K}_{rot}(\omega_s t)$ is shown in Section IV.A.

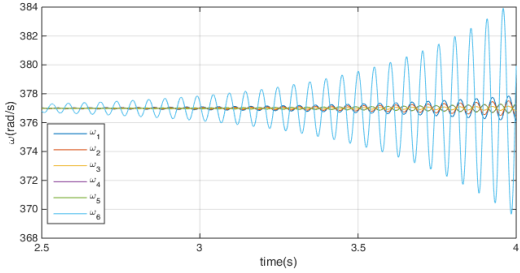
(a) rotating speed of each mass block

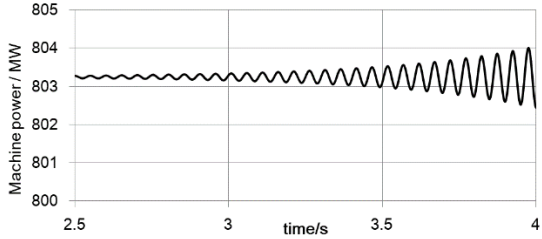
(b) mechanical power

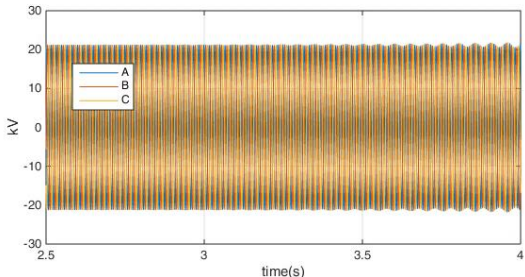
(c) transient voltage of generator

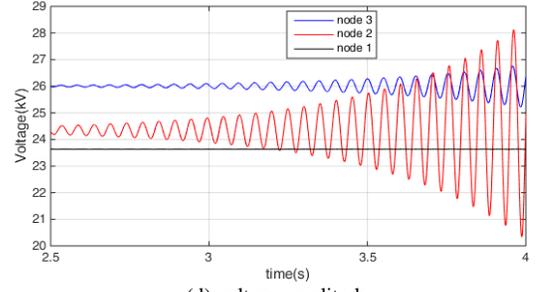
(d) voltage amplitude

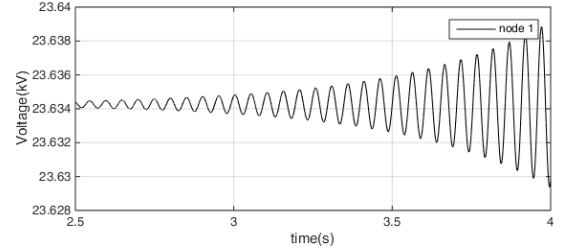
(e) voltage amplitude of node 1

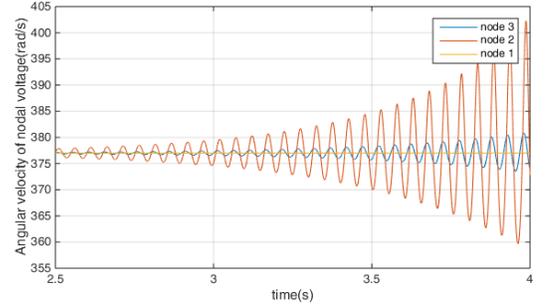
(f) angular velocity of rotating voltage vector

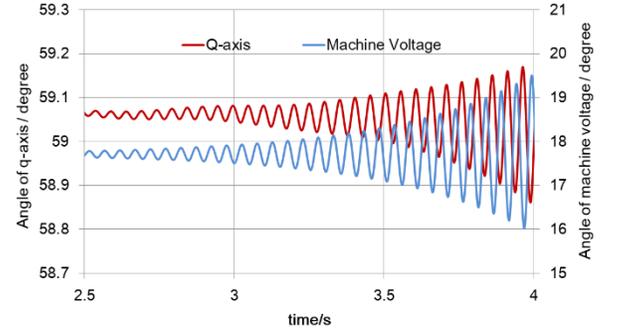
(g) angle of q-axis and phase angle generator voltage (in xy synchronous coordinate)

Fig. 6 Time domain results

Fig.6a shows the speed curves of the six mass blocks, which are all divergent oscillation at 19.64 Hz. It is assumed that the mechanical torques $T_1 \sim T_4$ in Fig.3 are all constant values. Due to the speed oscillation, the mechanical power (the product of torque and speed) would oscillate at 19.64 Hz, as shown in Fig.6b. Fig.6c shows the transient voltage curve of the generator, there is a divergent oscillation of 19.64Hz on the envelope of the voltage curve. In order to express this oscillation clearly, the AC voltage amplitude of three three-phase nodes could be obtained by using (42), as shown in Fig.6d. The voltage of node 1 is greatly affected by the infinite bus voltage, so the oscillation amplitude of node 1 is small. Fig.6e shows the voltage curve of node 1 detailed.

$$u_m = \sqrt{u_\alpha^2 + u_\beta^2} \quad (42)$$

Fig.6f shows the angular velocity of the rotating space voltage vector of the three AC nodes, which is calculated by (43).

$$\omega_u = \dot{\theta}_u = \frac{d\theta_u}{dt}, \quad \theta_u = \arctan\frac{u_\beta}{u_\alpha} \quad (43)$$

References [19] and [20] defined $\omega_u$ as the dynamic frequency of voltage. Since frequency is generally defined as the reciprocal of the period, and the definition of $\omega_u$ in (43) is the change rate of space angle, and the differential operation does not need a complete period to define $\omega_u$. Therefore, this paper names it rotating angular velocity rather than dynamic frequency.

In Fig.6d and Fig.6e, AC voltage curves show a divergent oscillation of 19.64 Hz around their respective rated voltages (26.00 kV, 24.36 kV and 23.63 kV). According to the definition of the stability of the equilibrium point shown in Fig.2, the voltages of three AC nodes are unstable. In Fig.6f, the angular velocity curves of the three AC voltages also show a divergent oscillation of 19.64 Hz around the rated angular velocity of 376.9911 rad/s. According to the definition of equilibrium point stability shown in Fig.2, the frequency of the three AC node voltages (the angular velocity of the voltage vector) is unstable. the rotor angle and generator voltage angle are also unstable, as shown in Fig.6a and Fig.6g.

We have come to the conclusion that traditional analysis of power angle stability, frequency stability, and voltage stability only discusses one or several elements in the state space vector of (40).

## VI. Comparing with Other Modeling Methods

### A. Comparing with EMTP Model

The traditional EMTP (Electro-magnetic Transient Program) software calculates and solves the synchronous machine access to the power grid according to the calculation block diagram shown in Fig.7 [21]. It divides the whole system into three parts: circuit, generator and shafting. Each part is calculated independently and solved alternately. It could not offer an unified differential equation for analytical analysis. In the EMTP software, circuit is described by the equivalent admittance matrix and the historical current source, which is an algebraic equation, not a differential equation.

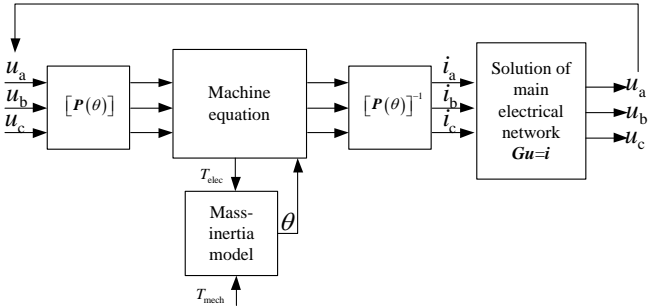

Fig.7 Calculation process of EMTP software

In this paper, the differential Equation (25) obtained by Lagrangian variational principle can be divided into several equations, as shown in (44)-(47).

Multi-mass block shafting equation

$$J\ddot{\theta} + D\dot{\theta} + K\theta = T + T_E \quad (44)$$

Generator equation with dq transformation

$$i_{machine} = \Gamma(\theta)\Psi_{machine} = \left[P(\theta)\Gamma_{dq}P^{-1}(\theta)\right]\Psi_{machine} \quad (45)$$

$$T_E = -\frac{1}{2}\Psi^T\Gamma'(\theta)\Psi \quad (46)$$

Power grid equation

$$K_C\ddot{\Psi} + K_R\dot{\Psi} + K_L\Psi = i_s - i_{machine} \quad (47)$$

When discretize (47) using the trapezoidal method, the obtained iterative equation is equivalent to the EMTP equation $Gu=i$ [14]. Therefore, (44)-(47) is equivalent to the calculation module in Fig.7, i.e., (25) provides a complete differential equation describing the electromagnetic transient process. Theoretically, since (25) is equivalent to the calculation model in EMTP software, both of them would get the same accurate time-domain results.

In fact, it is difficult to obtain accurate time-domain results of rotating generator with EMTP software. Larger numerical errors would submerge the oscillations caused by small disturbances. The numerical error of the EMTP model mainly comes from two aspects.

*1) EMTP method couldn't assign initial values to the system.*

The EMTP method establishes an algebraic equation of the form **Gu=i** for the electrical network [12][13], which is not a differential equation, so it couldn't directly assign initial values to the system. The synchronous generator model in PSCAD/EMTDC software also needs the initialization process of 1-2 seconds. Firstly, an ideal voltage source is used to replace the generator model, so that the electrical network can work stably at the specified equilibrium operating point. Then switch to the operating state of the generator model to simulate the transient process of the generator. In the system of Fig.3, the damping resistance is very small, which requires a very long initialization time to damp the oscillation caused by the initial value of the circuit, to ensure the system operating at the equilibrium operating point. As shown in Fig.8, when PSCAD software is used to simulate the system in Fig.3, the initialization time of 2s is not enough to make the voltage value of No.2 node converge to 24.36 kV required by the equilibrium operating point.

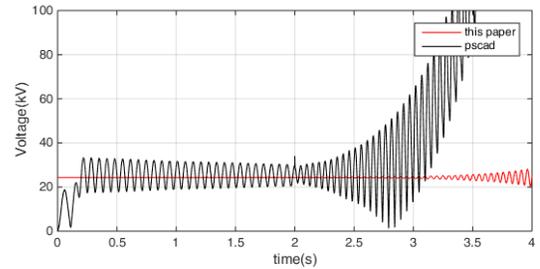

Fig.8 Voltage curve of node 2

Even in the example provided by PSCAD software (shown in Appendix C), when the ideal power supply is switched to the generator operation (1.5s), the three-phase voltage value on the series capacitor is still not damped by the resistance in the circuit to the balanced operating point, as shown in Fig.9. After 1.5 seconds, the voltage oscillation in the initialization process still does not disappear, and the new voltage oscillation is excited. The capacitor voltage curve is the superposition of two sets of oscillations.

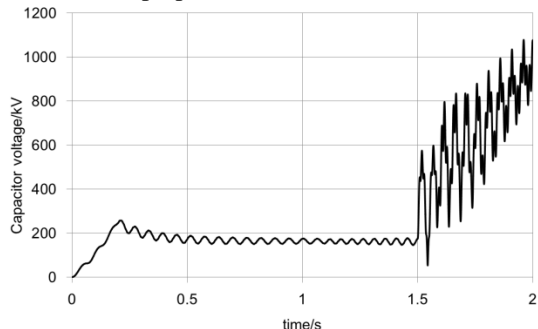

Fig.9 Simulation results of standard example in PSCAD

*2) time delay in module alternation calculation*

Since EMTP iterates calculate the mechanical shafting, generator equation and circuit equation in turn [21], there would be a time delay (equal to simulation time step) between each two modules which need to exchanging information. The time delay would bring large numerical error. In this paper, the dynamic model (25) is a complete differential equation. The complete equation is solved by numerical integration method in each simulation step, and there is no time delay. For example, a system shown in Fig.10a that does not contain capacitors, it could easily stabled at the steady-state operation point. The simulation result in 100μs time step, is shown in Fig.10b and Fig.10c.

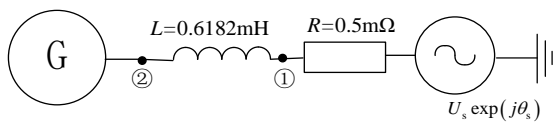

(a) Stable single machine system

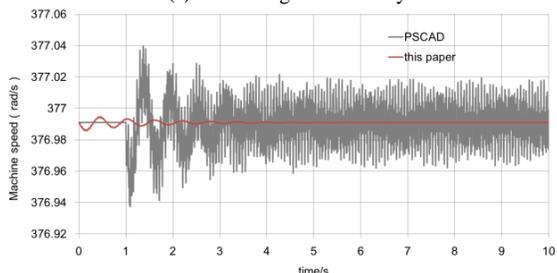

(b) Generator speed

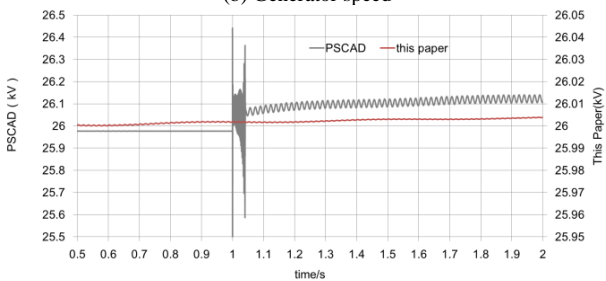

(c) Voltage of node 2

Fig.10 Simulation results of a stable system

In Fig. 10, the proposed dynamic modeling method presents a good result and converges rapidly, while the EMTP method (PSCAD/EMTDC) keeps a significant calculation error after the initialization process in 1 second.

*B. Comparing with the State Space Method*

Matlab/Simulink/SimPowerSystems uses state space method to model the circuit, the inductance current and capacitance voltage are selected to be the state variables. The mechanical shafting and generator equations still follow the calculation block diagram shown in Fig.7 to simulate the electromagnetic transient process.

In the circuit system, the generator is molded as current source, which is controlled by the generator equation. Current source The series connection of current source and inductor is ill-conditioned circuit [22]. So it is impossible to form the complete system state equation. It is because the state space method selects the current $I$ of the inductor branch as the state variable, and the change rate $dI/dt$ of the state variable is determined by the state equation. When the inductance branch is directly connected in series with the current source, the current value $I$ of the inductance branch and its change rate $dI/dt$ are determined by the output of the current source. It is logical contradiction. So, SimPowerSystems software will report an error.

However, the proposed dynamic modeling method and EMTP method could be applied to power systems with any connection mode.

*C. Comparing with Other Analytical Methods*

References [23] and [24] directly established a small disturbance differential equation in the form of $\frac{d\Delta x(t)}{dt} = A(t)\Delta x(t)$ in the xy synchronous coordinate system for the same system as Fig.3, which is used for modal analysis to determine the oscillation frequency. However, this small disturbance differential equation derived in the dq coordinate system couldn't be transformed into a complete differential equation $\frac{dx}{dt} = f(x,t)$ of the system in the stationary coordinate system, and couldn't be used to describe the complete electromagnetic transient oscillation divergence process. So [24] adopted the EMTP simulation results for verification. In [24], the differential equation used for analytical analysis and the mathematical model used for electromagnetic transient simulation are two different models, they couldn't be derived from each other.

According to the motion stability's definition, the intermediate function $d\Delta x/dt = g(\Delta x(t),t) = f(x(t)+\Delta x(t),t) - f(x,t)$ is required, to link the motion equation $dx/dt = f(x,t)$, and small disturbance equation $d\Delta x(t)/dt = A(t)\Delta x(t)$. Otherwise the derivation of mathematical equations couldn't be guaranteed correct. And this derivation process is one-way street: we could only obtain the small disturbance differential equation $d\Delta x/dt = g(\Delta x(t),t)$ from the motion equation $dx/dt = f(x,t)$, while the reverse direction does not work.

## D. Synchronous generator model in mechanics theory

The textbook of classical mechanics [26] gave an example of a circuit system. It selected the loop current $I = \dot{Q}$ as the generalized speed, and selected the loop charge $Q$ as the generalized coordinate, to obtain the dynamic equation. In [15], the node flux linkage was selected as the generalized coordinate to obtain the dynamic equation. They are two different ways. The analogy between circuit system and mechanical system is shown in Fig.11 and Table 4.

It should be noted that for dynamic modeling of synchronous generators, only the node flux could be selected as the generalized coordinate of the system. Otherwise, if the charge is chosen as the generalized coordinate, the magnetic field energy in the synchronous generator needs to be presented as $E_m(\theta,\dot{Q}) = \frac{1}{2}\dot{Q}^T L(\theta)\dot{Q}$. It is a function relay on both generalized coordinates $\theta$ and generalized velocity $\dot{Q}$. The energy function $E_m(\theta,\dot{Q})$ could not be simply classified into kinetic energy or potential energy, the Lagrangian function of the system couldn't be directly obtained.

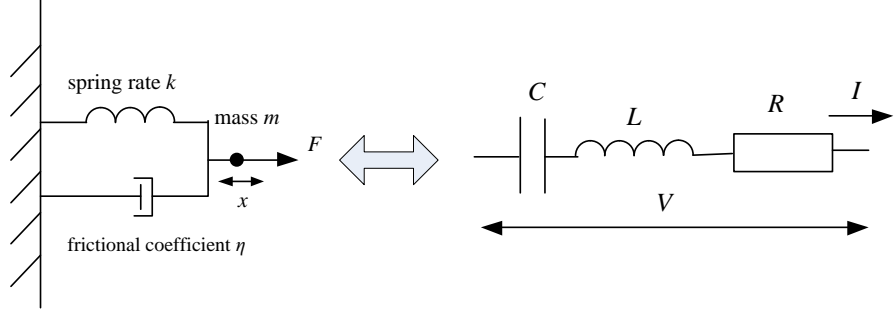

Fig. 11 Analogy of mechanical and electrical

Tab. 4 Analogy of mechanical system and RLC circuit system

| Variables or equations | Mechanical system | Circuitry | |
|---|---|---|---|
| | | Taking branch charge as generalized coordinate [26] | Taking the node flux linkage as the generalized coordinate |
| kinetic energy, potential energy, lagrangian function | $T = \frac{1}{2}m\dot{x}^2$, $U = \frac{1}{2}kx^2$ $\mathcal{L}(x,\dot{x}) = \frac{1}{2}m\dot{x}^2 - \frac{1}{2}kx^2$ | $T = \frac{L\dot{Q}^2}{2}$, $U = \frac{Q^2}{2C}$ $\mathcal{L}(x,\dot{x}) = \frac{L\dot{Q}^2}{2} - \frac{Q^2}{2C}$ | $T = \frac{1}{2}\dot{\Psi}^T K_C \dot{\Psi}$, $U = \frac{1}{2}\Psi^T K_L \Psi$ $\mathcal{L}(x,\dot{x}) = \frac{1}{2}\dot{\Psi}^T K_C \dot{\Psi} - \frac{1}{2}\Psi^T K_L \Psi$ |
| Rayleigh dissipation function | $\mathcal{R} = \frac{1}{2}\eta\dot{x}^2$ | $\mathcal{R} = \frac{1}{2}R\dot{Q}^2 = \frac{1}{2}RI^2$ | $\mathcal{R} = \frac{1}{2}\dot{\Psi}^T K_R \dot{\Psi}$ |
| Dynamical equation obtained by Lagrangian equation | $m\ddot{x} + \eta\dot{x} + kx = 0$ | $L\ddot{Q} + R\dot{Q} + \frac{Q}{C} = 0$ | $K_C\ddot{\Psi} + K_R\dot{\Psi} + K_L\Psi = 0$ |

## VII. Conclusions

In this paper, a typical power system with a synchronous generator containing a multi-mass mechanical shafting connected to the grid through LC series compensation is taken as an example. The dynamic model of the system is obtained based on the Lagrangian mechanics. The obtained dynamic model could be used for time-domain numerical simulation of electromagnetic transient, which breaks through the theoretical limitation that the electromagnetic transient model cannot be used for Lyapunov motion stability analysis. By selecting the phase angle of the ideal power supply as the reference angle, the differential equation of the dynamic model is transformed into the synchronous rotating coordinate system, and the motion stability problem is transformed into the equilibrium point stability problem.

The Lyapunov's first method is used to analyze the dynamic model in the synchronous coordinate system, and the frequency of divergent oscillation and its corresponding oscillation mode vector could be obtained. At the same time, the time domain numerical calculation of the dynamic model in the stationary coordinate system could obtain the accurate divergent oscillation curve. The model in the synchronous coordinate system and the model in the stationary coordinate system are the same dynamic model. In the dynamic model, the node voltage, the node frequency (the rotation angular velocity of the node voltage vector) and the generator angle (including the rotor angle and the terminal voltage phase angle) are one element or a combination of several elements in the generalized velocity vector and the generalized coordinate vector. This provides the feasibility for studying the unified stability of power systems.

# APPENDIX A: INDUCTANCE COEFFICIENT MATRIX OF SYNCHRONOUS GENERATOR

The generator inductance coefficient matrix in the dq coordinate system is:

$$\boldsymbol{\Gamma}_{dq} = \begin{bmatrix} L_d & 0 & L_{df} & L_{dD} & 0 & 0 \\ 0 & L_q & 0 & 0 & L_{qg} & L_{qQ} \\ L_{fd} & 0 & L_f & L_{fD} & 0 & 0 \\ L_{Dd} & 0 & L_{fD} & L_D & 0 & 0 \\ 0 & L_{gq} & 0 & 0 & L_g & L_{gQ} \\ 0 & L_{Qq} & 0 & 0 & L_{gQ} & L_Q \end{bmatrix}^{-1} \begin{bmatrix} \Gamma_d & 0 & \Gamma_{df} & \Gamma_{dD} & 0 & 0 \\ 0 & \Gamma_q & 0 & 0 & \Gamma_{qg} & \Gamma_{qQ} \\ \Gamma_{fd} & 0 & \Gamma_f & 0 & 0 & 0 \\ \Gamma_{Dd} & 0 & 0 & \Gamma_D & 0 & 0 \\ 0 & \Gamma_{gq} & 0 & 0 & \Gamma_g & \Gamma_{gQ} \\ 0 & \Gamma_{Qq} & 0 & 0 & \Gamma_{gQ} & \Gamma_Q \end{bmatrix}$$

The coefficient matrix $\boldsymbol{\Gamma}(\theta)$ is a function of $\theta$. $\boldsymbol{\Gamma}(\theta) = \boldsymbol{B}^{\mathrm{T}} \boldsymbol{\Gamma}_{\alpha\beta}(\theta) \boldsymbol{B}$, where, $\boldsymbol{B}$ is the incidence matrix.

$$\boldsymbol{\Gamma}_{\alpha\beta}(\theta) = \left[ \boldsymbol{P}(\theta) \boldsymbol{\Gamma}_{dq} \boldsymbol{P}^{-1}(\theta) \right]$$

$$= \begin{bmatrix} \Gamma_d \cos^2\theta + \Gamma_q \sin^2\theta & (\Gamma_d - \Gamma_q)\sin\theta\cos\theta & \Gamma_{df}\cos\theta & \Gamma_{Dd}\cos\theta & -\Gamma_{gq}\sin\theta & -\Gamma_{Qq}\sin\theta \\ (\Gamma_d - \Gamma_q)\sin\theta\cos\theta & \Gamma_q\cos^2\theta + \Gamma_d\sin^2\theta & \Gamma_{df}\sin\theta & \Gamma_{Dd}\sin\theta & \Gamma_{gq}\cos\theta & \Gamma_{Qq}\cos\theta \\ \Gamma_{df}\cos\theta & \Gamma_{df}\sin\theta & \Gamma_f & \Gamma_{fD} & 0 & 0 \\ \Gamma_{Dd}\cos\theta & \Gamma_{Dd}\sin\theta & \Gamma_{fD} & \Gamma_D & 0 & 0 \\ -\Gamma_{gq}\sin\theta & \Gamma_{gq}\cos\theta & 0 & 0 & \Gamma_g & \Gamma_{gQ} \\ -\Gamma_{Qq}\sin\theta & \Gamma_{Qq}\cos\theta & 0 & 0 & \Gamma_{gQ} & \Gamma_Q \end{bmatrix}$$

$$\boldsymbol{\Gamma}'(\theta) = \boldsymbol{B}^{\mathrm{T}} \boldsymbol{\Gamma}'_{\alpha\beta}(\theta) \boldsymbol{B},$$

$$\boldsymbol{\Gamma}'_{\alpha\beta}(\theta) = \frac{\partial \boldsymbol{\Gamma}_{\alpha\beta}(\theta)}{\partial \theta}$$

$$= \begin{bmatrix} -(\Gamma_d - \Gamma_q)\sin 2\theta & (\Gamma_d - \Gamma_q)\cos 2\theta & -\Gamma_{df}\sin\theta & -\Gamma_{Dd}\sin\theta & -\Gamma_{gq}\cos\theta & -\Gamma_{Qq}\cos\theta \\ (\Gamma_d - \Gamma_q)\cos 2\theta & (\Gamma_d - \Gamma_q)\sin 2\theta & \Gamma_{df}\cos\theta & \Gamma_{Dd}\cos\theta & -\Gamma_{gq}\sin\theta & -\Gamma_{Qq}\sin\theta \\ -\Gamma_{df}\sin\theta & \Gamma_{df}\cos\theta & 0 & 0 & 0 & 0 \\ -\Gamma_{Dd}\sin\theta & \Gamma_{df}\cos\theta & 0 & 0 & 0 & 0 \\ -\Gamma_{gq}\cos\theta & -\Gamma_{gq}\sin\theta & 0 & 0 & 0 & 0 \\ -\Gamma_{Qq}\cos\theta & -\Gamma_{Qq}\sin\theta & 0 & 0 & 0 & 0 \end{bmatrix}$$

The small disturbance of the electromagnetic torque needs to consider the second-order matrix derivative $\boldsymbol{\Gamma}''(\theta) = \boldsymbol{B}^{\mathrm{T}} \boldsymbol{\Gamma}''_{\alpha\beta}(\theta) \boldsymbol{B}$ ( see (39) in the text ), where

$$\boldsymbol{\Gamma}''_{\alpha\beta}(\theta) = \frac{\partial^2 \boldsymbol{\Gamma}_{\alpha\beta}(\theta)}{\partial \theta^2}$$

$$= \begin{bmatrix} -2(\Gamma_d - \Gamma_q)\cos 2\theta & -2(\Gamma_d - \Gamma_q)\sin 2\theta & -\Gamma_{df}\cos\theta & -\Gamma_{Dd}\cos\theta & \Gamma_{gq}\sin\theta & \Gamma_{Qq}\sin\theta \\ -2(\Gamma_d - \Gamma_q)\sin 2\theta & 2(\Gamma_d - \Gamma_q)\cos 2\theta & -\Gamma_{df}\sin\theta & -\Gamma_{Dd}\sin\theta & -\Gamma_{gq}\cos\theta & -\Gamma_{Qq}\cos\theta \\ -\Gamma_{df}\cos\theta & -\Gamma_{df}\sin\theta & 0 & 0 & 0 & 0 \\ -\Gamma_{Dd}\cos\theta & -\Gamma_{Dd}\sin\theta & 0 & 0 & 0 & 0 \\ \Gamma_{gq}\sin\theta & -\Gamma_{gq}\cos\theta & 0 & 0 & 0 & 0 \\ \Gamma_{Qq}\sin\theta & -\Gamma_{Qq}\cos\theta & 0 & 0 & 0 & 0 \end{bmatrix}$$